\documentclass[preprint,aps]{revtex4}

\begin{document}
\title{A decoy-state protocol for quantum cryptography with 4 
intensities of coherent light}
\author{Xiang-Bin Wang\thanks{Email address: wang@qci.jst.go.jp}\\
IMAI Quantum Computation and Information Project,
ERATO, JST,
Daini Hongo White Bldg. 201, \\5-28-3, Hongo, Bunkyo,
Tokyo 133-0033, Japan}
\begin{abstract} 
In order to beat any type of 
photon-number-splitting attack, we
 propose a protocol for quantum key distributoin (QKD) using
4 different intensities of pulses. They are vacuum and coherent states with
mean photon number $\mu,\mu'$ and $\mu_s$. 
$\mu_s$ is around 0.55 and this class of pulses are used
 as the main signal states. The other two classes of
coherent states ($\mu,\mu'$) are also used signal states but
their counting rates should be studied jointly with the vacuum.
We have shown that, given the typical set-up in practice,
the key rate from the main signal pulses is quite close to  
the theoretically allowed maximal rate in the case
given the small overall transmittance of $10^{-4}$.  
\end{abstract}
\maketitle
\section{Introduction}
Quantum key distribution(QKD) has drawn much attentions from scientists.
Different from the classical cryptography, quantum
key distribution(QKD)\cite{wies,gisin,bene} can help
two remote parties to set up the 
secure key by non-cloning theorem\cite{woot}.
Further, proofs for the unconditional security over
noisy channel have been given\cite{shor2,lo3,maye,ekert}.
The security of practical QKD with weak coherent states has also been shown\cite{inl,gllp}. 
However there are still some limitations for QKD in practice, especially over
long distance.
 In particular,
large loss of channel seems to be the main challenge to the long-distance
QKD with weak coherent states.
A dephased coherent state $|\mu e^{i\theta}\rangle$
is actually a mixed state of
\begin{eqnarray}
\rho_\mu=\frac{1}{2\pi}
\int_0^{2\pi}|\mu e^{i\theta}\rangle\langle\mu e^{i\theta}|{\rm d}\theta
=\sum_n P_n(\mu)|n\rangle\langle n|\label{coherent0}
\end{eqnarray}   
and 
$
P_n(\mu)=\frac{\mu^ne^{-\mu}}{n!}.
$
Here
$\mu$ is a non-negative number.
In practice, especially in doing long-distance QKD, the channel
transmittance $\eta$ can be rather small. If 
$\eta<1-e^{-\mu}-\mu e^{-\mu}$, Eavesdropper (Eve) in principle
can have the full information of Bob's sifted key by the photon-number-splitting
(PNS) attack\cite{bra}:
Eve blocks all single-photon pulses and part of multi-photon pulses and 
separates each of the remained multi-photon pulses into two parts therefore each part contains at least one
photon. She keeps one part and sends the other part to Bob, through a 
lossless channel. 

If the channel is not so lossy, Alice and Bob can still
set-up the unconditionally secure final key with a key rate\cite{gllp} 
\begin{eqnarray}
r=1-\Delta-H(t)-(1-\Delta)H(t/(1-\Delta))
\label{dllp}
\end{eqnarray}
if we use a random classical CSS code\cite{shor2} to distill the final key\cite{gllp}. 
Here $t$ is the flipping error rate, $\Delta$ is the fraction of tagged signals\cite{gllp},
i.e. the fraction for those counts in cases when
 Alice sends out a multi-photon pulse. 
The functional $H(x)=-x\log_2x-(1-x)\log_2(1-x)$.
From the above formula we see that a tight
  bound for $\Delta$ is rather important in both
key rate and the threshold of flipping rates.

It is possible to use single-photon source\cite{single} in
the next generation of practical QKD after the technique is fully matured, but it seems 
not likely  in the near future. Moreover, it seems not to be the
best choice from economic viewpoint. There are at least two realistic methods
so far: strong-reference-light\cite{srl} method and decoy-state method\cite{hwang}.

Originally, the PNS attack has been investigated where Alice and Bob 
monitor only how many non-vacuum signals arise, and how many errors 
happen. However, it was then shown\cite{kens1} that the simple-minded method
does not guarantee the final security. It is shown\cite{kens1}  that in a 
typical parameter regime nothing changes if one starts to monitor the 
photon number statistics as Eve can adapt her strategy to reshape the 
photon number distribution such that it becomes Poissonian again. 
A very important method for was then proposed by Hwang\cite{hwang}, where a method
for {\it unconditional} verification of the multi-photon counting rate (MPCR) is given.
Using Hwang's result, one can faithfully estimate the upper bound
of $\Delta$ through  decoy-pulses, given $whatever$ type of PNS attack.
The value of upper bound estimated there is much decreased than that in
worst-case estimation. However, Hwang's method does not produce
a sufficiently tight bound, though it is an unconditional verification. 
 For example, in the case of $\mu=0.3$,
by Hwang's method,
the the optimized verified upper bound of $\Delta$ is $60.4\%$.
With the value $\Delta=60.4\%$. As it was mentioned\cite{hwang,tot}, one can
combine the decoy-state method with GLLP\cite{gllp}.  By eq(\ref{dllp}),
the key rate must be low in practice if we use the the bound value given by Hwang\cite{hwang}.
Latter, Lo and co-workers studied the subject\cite{tot,lo4}.  
However, their main protocol\cite{tot} 
seems to be inefficient in practice, because it requires
infinite number of classes of different coherent states to work as the decoy states. 
A detailed version of their main protocol has been presented recently\cite{lolo}.
Prior to Ref.\cite{tot}, a simple idea of using vacuum and $very$ weak cohernet states
as decoy states was shortly stated\cite{lo4}:``On one hand, by using a vacuum as decoy state, 
Alice and Bob can verify the so called dark count rates of their detectors. On the other
hand, by using a very weak coherent pulse as decoy state, Alice and Bob can easily lower bound
the yield (transmittance) of single-photon pulses.'' This idea obviously
works in the ideal case with infinite pulses\cite{tot}. However, 
as we shall show it latter, given a very lossy channel and finite number
of pulses, the total number of counts of those very weak coherent states
can be too small to be useful for a faithful stastical estimation. 
 \section{Our protocol and results}
Recently, the author proposed an efficient decoy-state protocol\cite{my}
with vacuum and two coherent states of $\mu,\mu'$ which are used for
both decoy and signal.
Here, we propose a modified protocol which further improves the key rate.
In the modified protocol, coherent states with average photon number
$\mu_s$ is used for the main signal state. 
Coherent states with average photon number $\mu$, $\mu'$ are used
for both signal and decoy states( i.e., they are used as signal states but
their counting rates are also observed and used in the protocol.) 
Vacuum is used only for testing.
The main idea of this work is: According to the transmittance of the
physical channel, we first choose a reasonable
value for $\mu$, e.g. 0.1 or 0.22 and then find a good value $\mu'$
so that $\mu$ and $\mu'$ will help to verify a satisfactorily
value of transmittance of single-photon pulses, $s_1$. According to
$s_1$, we then choose the value $\mu_s$ so that the key rate of
main signal states is maximized. In a real protocol, Alice is supposed to calculate
these values according to the transmittance of physical channel {\it in advance}.
 Alice mixes all classes of 
pulses and sends them to Bob and then verify the 
value of the single-photon transmittance according to the counting rates of
states of vacuum, $\mu$ and $\mu'$. If the verified value
is too much smaller than the expected value, they give up the protocol.   
Our protocol has the following properties:
(1), The protocol uses only 4 classes of states. Except for vacuum, all 
pulses have the reasonable intensity and all of them can be used as signal
states.
(2), The protocol gives a key rate ranges from $77\%$ to $88\%$
of that of the theoretically allowed key rate, given the
overall transmittance of $10^{-4}$ or $10^{-3}$. 
(3), The protocol assumes typical real-world set-ups of  QKD in practice therefore
it applies for {\it real-world} protocols with coherent states.
Let's start from an estimation of the theoretically allowed maximum key rate
(TAMKR)
with coherent states.
\subsection{theoretically allowed maximum key rate}
To see the TAMKR, we consider an ideal protocol:
\\{\it Ideal protocol:} Alice and Bob exactly 
uses $N_{s}$ single-photon pulses to test the 
transmittance and quantum bit error rate(QBER) of
 all single-photon pulses. 
The dark count is zero and the channel transmittance
is $\eta$.
They use coherent states to generate the key. 
Suppose the tested QBER is $t'_1$ and then they
can upper-bound the QBER of those single-photon states
in signal pulses by 
\begin{eqnarray}
t_1\le (1+\delta) t'_1.\label{QBER}
\end{eqnarray}  
 They use coherent state
with intensity $\mu$ to generate the key. 
According to eq.(\ref{dllp}), the overall key rate is
\begin{eqnarray}
R=\eta \mu [1-2H(t_1)-(1- e^{-\mu})(1-H(t_1))].
\end{eqnarray}
They may choose an appropriate value $\mu$ to maximize $R$.
For example, given $t_1=0$, maximized value is
$R=\eta\mu e^{-\mu}$ at the point of $\mu=1$.
In this papeer, we shall consider the typical case that the QBER is
$t_1=0.03$ and for this value the TAMKR is
\begin{eqnarray}
R_{TAMKR}=0.149\eta
\end{eqnarray}
with $\mu=0.572$.
\subsection{elementary results}
In our protocol, the $BB84$ or other quantum-bit states are encoded in 
each coherent pulses
(except for vacuum pulses.) What we shall study is {\it not} the $BB84$ state
or other qubit state for cryptography itself, 
we shall only study how to overcome the PNS attack.  
Alice switch the intensity (mean photon number) 
of each pulse
randomly among 4 values,  $0,\mu,\mu',\mu_s$. (These values have nothing to do
with BB84-state preparation. Except for vacuum, each pulses caries a state randomly
chosen from BB84-set and there is no relation between intessity and the carried BB84-state
in any pulse.)
We first use the pulses with intensities of  $\mu,\mu'$ to
estimate a lower bound on the overall transmittance of single
photon pulses and then calculate the key rate of the main signal states
by this lower bound.
For simplicity, we denote those pulses produced in state  
$|\mu_s e^{i\theta}\rangle, |\mu e^{i\theta}\rangle, |\mu' e^{i\theta}\rangle,|0\rangle$ as
class $Y_s,Y_\mu, Y_{\mu'}$ and $Y_0$, respectively.
In the protocol $\theta$
is randomized. They observe the counting rates of each classes
so we regard $s_0,S_\mu,S_{\mu'},S_{\mu_s}$ as {\it known} parameters
and  notations $s_0,S_\mu,S_{\mu'},S_{\mu_s}$ are counting rates for
pulses in classes of  $Y_0,Y_\mu, Y_{\mu'},Y_{\mu_s}$, respectively.
They
verify the lower bound of
single photon transmittance $s_1$ using the measured values of 
$s_0,S_\mu,S_{\mu'}$.
With $s_1$ being verified, they can distill the final key from 
all classes of pulses except for $Y_0$.
Given the transmittance, not all values of $\mu,\mu'$ will work
same effectively.
They should choose appropriate values of $\mu,\mu'$ so that they can verify
a large lower bound of $s_1$. They should also choose an appropriate
$\mu_s$ so that the key rate of pulses in this class is maximized.
That is to say, there are two steps of optimization. First they need
good values of $\mu,\mu'$ to verify lower bound of $s_1$ tightly.
Second, given $s_1$, normally, neither $\mu$ nor $\mu'$ maximizes the
key rate, they need to use another intensity of states, $\mu_s$
as their main signal pulses. 
If there is no Eve or Eve hides her presence, after the protocol
they must be able to verify everything as expected, and they can indeed
obtain satisfactory results. If the verified results about $s_1$ is too much
larger than what was expected, they give up the protocol. In this paper,
the calculation for choosing $\mu,\mu'$ is similar to my previous
work\cite{my}, but we show something more: after adding another class of
coherent pulses $Y_s$,
the key rate of that class of pulses is approaching 
the theoretically allowed value. 
 
We first define the {\it counting rate} of {\it any} state $\rho$:
 the probability that Bob's detector clicks whenever
a state $\rho$ is {\it sent out} by Alice.
We $disregard$ what state Bob may receive here. This {\it counting rate}
is called as the {\it yield} in other literatures\cite{hwang,tot}.
 For convenience, we $always$ assume 
\begin{eqnarray}
\mu'>\mu; \mu' e^{-\mu'} > \mu e^{-\mu} \label{condition}
\end{eqnarray} in this paper.
 Alice is the only person who knows which pulse belongs to which class.
After received all pulses from Alice, Bob announces which pulse has caused a click and
which pulse has not. 
At this stage, Alice has already known the {\it counting rates} 
of pulses in each of the four classes, $\{Y_0,Y_\mu,Y_{\mu'},Y_s\}$.
Their task is to verify the lower bound of $s_1$, or equivalently,
the upper bound of $\Delta$, the fraction of multi-photon counts among all counts caused
by pulses in  class $Y_\mu$.

 A dephased coherent state $|\mu e^{i\theta}\rangle$ 
has the following convex form: 
\begin{eqnarray}
\rho_{\mu}= e^{-\mu}|0\rangle\langle0|+\mu e^{-\mu}|1\rangle\langle 1|
+c\rho_c\label{oo}
\end{eqnarray}
and $c=1-e^{-\mu}-\mu e^{-\mu}>0$, 
\begin{eqnarray}
\rho_c=\frac{1}{c}\sum_{n=2}^\infty P_n(\mu)|n\rangle\langle n|\label{coherent}.
\end{eqnarray}
Similarly, state  $|\mu' e^{i\theta}\rangle$ after dephasing is
\begin{eqnarray}
\rho_{\mu'}=e^{-\mu'}|0\rangle\langle0|+\mu' e^{-\mu'}|1\rangle\langle 1|
+c\frac{\mu'^2 e^{-\mu'}}{\mu^2 e^{-\mu}}\rho_c + d\rho_d\label{dd}
\end{eqnarray}
and $d=1-e^{-\mu'}-\mu' e^{-\mu'}-c\frac{\mu'^2 e^{-\mu'}}{u^2 e^{-\mu}} \ge 0$.
$\rho_d$ is a density operator. (We shall only use the fact that $d$ is non-negative and
$\rho_d$ $is$ a density operator.) In deriving the above convex form, we have used the fact
$P_n(\mu')/P_2(\mu')> P_n(\mu)/P_2(\mu)$ for all $n>2$, given the conditions of eq.(\ref{condition}).
With these convex forms of density operators,
it is equivalent to say that Alice sometimes sends nothing 
($|0\rangle\langle 0|$), sometimes sends $|1\rangle\langle 1|$,
sometimes sends $\rho_c$, sometimes sends $\rho_d$ and so on,  
though Alice does not know which time she has sent out which one of these states. In each individual sending, 
she only knows which class the sent state belongs to.
We shall use notations $s_0,S_\mu,S_{\mu'},S_{\mu_s},s_1,s_{c},s_d$ for the 
{\it counting rates} of pulses in class $Y_0,Y_\mu,Y_{\mu'},Y_s$,
pulses in single-photon state, pulses in state $\rho_c$ and pulses in
state $\rho_d$,
respectively. 
Our goal is simply to find a formula relating 
$s_1$ or 
$\Delta$ with the quantities of $s_0,S_\mu,S_{\mu'}$ which are known to 
Alice and Bob already.
 Given any state $\rho$, nobody but Alice can tell whether it is from
class $Y_\mu$ or $Y_{\mu'}$. Asymptotically, we have  
\begin{eqnarray}
s_\rho(\mu)= s_\rho(\mu')
\end{eqnarray} 
and $s_\rho(\mu),s_\rho(\mu')$ are {\it counting rates} 
for state $\rho$ from class $Y_\mu$ and class 
$Y_{\mu'}$, respectively. 

The coherent state $\rho_{\mu'}$ 
is convexed by $\rho_c$ and other states. Given the condition
of eq.(\ref{condition}),  the probability of
$\rho_c$ in state $\rho_{\mu'}$ is larger than that in $\rho_{\mu}$. 
Therefore we can make a preliminary estimation of $s_c$. 
From eq.(\ref{dd}) we immediately obtain 
\begin{eqnarray}
S_{\mu'}= e^{-\mu'}s_0 + \mu' e^{-\mu'}s_1 + c\frac{\mu'^2 e^{-\mu'}}
{\mu^2 e^{-\mu}}s_c +ds_d.\label{origin}
  \end{eqnarray}
$s_0$ is known, $s_1$ and $s_d$ are unknown, but they can never be less than 0. Therefore
we have
\begin{eqnarray}
 e^{-\mu'}s_0 + \mu' e^{-\mu'}s_1 + c\frac{\mu'^2 e^{-\mu'}}{\mu^2 e^{-\mu}}s_c\le S_{\mu'}.
\label{crude00}\end{eqnarray}
From eq.(\ref{oo}) we also have
\begin{eqnarray}
e^{-\mu}s_0 + \mu e^{-\mu}s_1 + c s_c=S_{\mu}. \label{crude1}
\end{eqnarray}
Solving the above two constraints self-consistantly we have
\begin{eqnarray}
\Delta=\frac{cS_c}{S_\mu} \le
\frac{\mu}{\mu'-\mu}\left(\frac{\mu e^{-\mu} S_{\mu'}}{\mu' e^{-\mu'} S_{\mu}}-1\right) 
+\frac{\mu e^{-\mu}s_0 }{ \mu' S_\mu }\nonumber\\
s_1=\frac{1-\Delta-e^{-\mu}s_0/S_\mu}{\mu}e^{\mu}S_{\mu}.
\label{assym}
\end{eqnarray}
In particular, in the case $\eta<<1$ and there is no Eve., Alice and Bob must be able to 
verify the following facts:
\begin{eqnarray}
s_1=e^\mu (1-\Delta)\eta + [(1-\Delta)e^\mu-1]s_0/\mu
\end{eqnarray}
and, if we set $\mu'-\mu\rightarrow 0$ we have
\begin{eqnarray}
\Delta = \left. \frac{\mu \left(e^{\mu'-\mu}-1\right)}{\mu'-\mu}\right|_{\mu'-\mu\rightarrow 0}=\mu\label{mcount}
\end{eqnarray} 
in the protocol. (In eq.(\ref{mcount}) we have set $s_0=0$ for the
clarity of the main issue.  
This is close to the real value in the case of normal lossy channel, which is $1-e^{-\mu}$, given that $\eta<<1$.
From the above observation we can summarize two points: (1), Assymptotically, $\mu,\mu'$ should be chosen
 close to each other so as to  obtain a tight lower bound for $s_1$. (2), 
The over estimation of $\Delta$ by our protocol is  $\mu-(1-e^{-\mu})=\mu^2/2$. Therefore, the smaller $\mu$ is chosen,
the tighter our verification of  $\Delta,s_1$  is.
   However, we can not choose to set 
$\mu$ or $\mu'-\mu$ to be unlimittedly small in practice,
otherwise the protocol is neither stable nor secure due to the statistical 
fluctuation.
The results above are only for the asymptotic case. 
In practice, the number of pulses are always finite and negative effects from possible 
statistical fluctuation have to be considered. Otherwise, the protocol is {\it insecure}.
  Before going into  details of such a task, we give an example
 to see why the fluctuation can cause serious security problem if it is
disregarded. Consider a toy protocol: Alice and Bob use
 single-photon state as the decoy state
to test $s_1$ and use normal coherent state for key distillation.
 Suppose the total number of pulses of decoy states is
$10^{5}$ and they find 20 clicks at Bob's side for all decoy pulses. If they conclude that
$s_1= 2\times 10^{-4}$ the protocol is very insecure: there is substentially non-negligible
 probability that the real value of $s_1$ for signal pulses
is  only a half of that. Similar problem also occurs in 
the idea of Ref.\cite{lo4} where
{\it very} weak coherent state\cite{lo4} is used to replace the single-photon 
decoy state. 
Obviously, to lower bound the value of
$s_1$ by observing the counting rate of the very weak coherent states, 
the mean photon number $\mu_v$ of the very weak coherent states must be less than $\eta$.
Suppose we use $10^{10}$ pulses for the decoy pulses of $\mu_v$, then in average the total
counts of the very weak coherent states is $\mu\eta<\eta^2$. Given channel transmittance
of $\eta=10^{-4}$, in average, there would be only less than 100 counts for the decoy pulses 
of very weak coherent
states. This seems insufficient to make a faithful stastical estimation for the true value of $s_1$.  Moreover, in practice, the dark counts will make things 
even worse. Suppose they use a number of vacuum states class $Y_0$ to test the
dark count rate. 
Suppose in class $Y_0$, the tested dark count rate is $s_0=10^{-6}$.
Suppose there are $10^{10}$ pulses in class of very weak coherent state.
{\it In average}, they should find 10100 counts for this class, 10000 dark
counts and 100 counts caused by the very weak coherent states. However, this
is only the values {\it in average}. In a specific realization, they
will have no way to verify anything.
For example, if they observe 10100 counts for among $10^{10}$ pulses of very weak coherent states,
there is a non-negligible probability that
{\it all these counts are due to  the dark counts only} while the single-photon
state counting rate is actually zero. The idea\cite{lo4} of using very weak coherent state
as decoy state will require an unreasonably large number of pulses for a faithful statistical estimation\cite{comment}.
Besides the issue of statistical fluctuation, exactly producing the expected very weak coherent states itself
can be technically difficult.
Now we show how our protocol works in practical set-ups.
\subsection{numerical results of the protocol}
In practice, our task is stated as this: to verify a tight
 lower bound
of $s_1$ and the probability that the real value of $s_1$
for signal pulses in any class being
less than the verified lower bound 
is exponentially close to 0.

The counting rate of any state $\rho$ in class $Y_{\mu'}$
now can be slightly different from the counting rate of the same state $\rho$ 
from another class, $Y_\mu$, with non-negligible
probability. We shall use the primed
notation for the counting rate for any state in class $Y_{\mu'}$ and the original notation
for the counting rate for any state in class $Y_\mu$. 
Explicitly, eq.(\ref{crude00},\ref{crude1})
are now converted to
\begin{eqnarray}
  \left\{ \begin{array}{l} 
e^{-\mu}s_0 + \mu e^{-\mu}s_1 + c s_c=S_{\mu},
 \\ cs'_{c}\le \frac{\mu^2e^{-\mu}}{\mu'^2e^{-\mu'}}\left(S_{\mu'}
- \mu' e^{-\mu'} s'_1
- e^{-\mu'}s'_0\right) .
  \end{array}
  \right. \label{couple}
 \end{eqnarray}
Setting $s_x' = (1-r_x)s_x$ for $x=1,c$  and $s'_0=(1+r_0)s_0$ we obtain
\begin{eqnarray}
\mu' e^\mu \left[(1-r_c)\frac{\mu'}{\mu}-1\right]\Delta \le \mu e^{\mu'}S_{\mu'}/S_\mu
-\mu'e^{\mu}+[(\mu'-\mu)s_0+r_1s_1+r_0s_0]/S_\mu.
\end{eqnarray}
In the left side, if $\mu'$ and $\mu$ are too close, the factor
of $\Delta$ is very small. In the right side,  if $\mu'-\mu$ is too small,
term $r_1s_1$ will contribute effectively. Therefore, in practice, $\mu'$
and $\mu$ have to be a bit different.
The important question here is whether there are reasonable
values for $\mu',\mu$ so that our protocol can verify
a tight lower bound of $s_1$ even though the number of pulses is finite.
 The answer is yes.
Now the problem is actually this: given the normal case that they have found
$S_\mu=\eta\mu, S_{\mu'}=\eta\mu'$, (i.e., there is no Eve.), how tightly
they can lower bound $s_1$.
Given $N_1+N_2$ copies of state $\rho$,  suppose
the counting rate
for $N_1$ randomly chosen states is $s_{\rho}$ and the counting rate
for the remained states  is $s'_{\rho}$ the probability that $s_\rho-s'_\rho>\delta_\rho$
is
less than $\exp\left(-\frac{1}{4}{\delta_\rho}^2N_0/s_\rho\right)$
and $N_0 ={\rm Min}(N_1,N_2)$. Now we consider the difference of counting rates
for the same state from different classes, $Y_\mu$ and $Y_{\mu'}$.
 To make a faithful estimation 
for exponentially sure, we require 
${\delta_\rho}^2N_0/s_\rho =100$. This causes a relative fluctuation   
\begin{eqnarray}
r_\rho=\frac{\delta_\rho}{s_{\rho}}\le 10\sqrt{\frac{1}{s_{\rho}N_0}}\label{statis}.
\end{eqnarray} 
The probability of violation is less than $e^{-25}$.
 To formulate the relative fluctuation $r_1,r_c$
by $s_c$ and  $s_1$, we only need check the number of pulses in state $\rho_c$,
$|1\rangle\langle 1|$  in each classes in the protocol.
That is, using eq.(\ref{statis}),
we can replace $r_1,r_c$ in eq.(\ref{couple}) 
by $10e^{\mu/2}\sqrt{\frac{1}{\mu s_1N}}$, 
$10\sqrt{\frac{1}{c s_cN}}$, respectively
and $N$ is the number of pulses in class $Y_\mu$.
From this we can also see that value $\mu$ itself cannot be set too small,
otherwise the total number of single-photon pulses is too small therefore the
fluctuation is severe.
Since we assume the case where vacuum-counting rate is much less than
$S_\mu$, we shall omit the effect of fluctuation
in vacuum counting, i.e., we set $r_0=0$.
 With these inputs, eq.(\ref{couple}) can now be solved
numerically.
 The verified bound values of $s_1$ are listed in the following table I. 
 They are values that can be verified in the case that there is no Eve
(or Eve hides her presence). 
In obtaining those values, we first choose a 
reasonable value for $\mu$. According to $\mu$, we choose
an appropriate $\mu'$ therefore a tight bound for $s_1$ is  obtained.
\begin{table}
\caption{Verification of transmittance of single-photon pulse.
We need the pulses in class $Y_0,Y_\mu,Y_{\mu'}$ for verification. Class
 ${Y_\mu}$ or $Y_{\mu'}$ need $10^{10}$ pulses and $Y_0$ needs $2\times 10^{9}$. }
\begin{tabular}{c|c|c|c|c}
$\eta$ & $10^{-3}$ &$ 10^{-3}$ &$10^{-4}$ & $10^{-4}$ 
\\ 
\hline 
$s_0$ & $10^{-6}$ & $2\times 10^{-7}$&   $10^{-6}$ & $2\times 10^{-7}$\\
\hline
$\mu$ & 0.1 & 0.1& 0.22 & 0.1
\\
\hline
$\mu'$ & 0.27 & 0.26 & 0.48 & 0.35
\\ \hline
$s_1/\eta$ & 0.958 & 0.969 & 0.821 & 0.922
\end{tabular}
\end{table} 
Next, we shall consider the QBER. For a fair comparison of the ideal protocol
and our protocol, we assume the same channel and the same device for both
protocol. Therefore the bit-flip part should be equal. The bound of
phase-flip rate of our protocol should be larger than that in the ideal 
protocol, because here we have to assume all phase-flip
errors have happened to the single-photon pulses.
If the QBER in ideal protocol is $E$, then the phase-flip in our protocol is
\begin{eqnarray}
E' \le f E; f=  e^{\mu}\eta/s_1. 
\end{eqnarray}
Using eq.(\ref{dllp}) we have the formula for key rate on class $Y_s$:
\begin{eqnarray}
R_s=S_{\mu_s}[1-H(E)-H(fE)-\Delta_s (1-H(f E))].\label{rate11}
\end{eqnarray}
and $S_{\mu_s}$ is verified to be $\eta\mu_s$, $E$ is the meassured
error rate of the main signal pulses,
\begin{eqnarray}
\Delta_s=1- \frac{s_1(\mu_s)}{\eta} e^{-\mu_s}.
\end{eqnarray}
We shall assume that $E$ is bounded by $3\%$, the same with
$t_1$ in the {\it ideal protocol}. 
The key rates for class $Y_s$ in various cases is listed in table II. 
\begin{table}
\caption{ Final key rate.
The last raw is the ratio of key rate from main signal pulses
and the theoretically allowed maximal value. 
We have assumed the QBER for signal states in the {\it Ideal protocol}
is bounded by $t=3\%$. The number of pulses
of in $Y_s$ can be any number larger than $10^{10}$.}
\begin{tabular}{c|c|c|c|c}
$\eta,s_0$ & $10^{-3},10^{-6}$ &$ 10^{-3},2\times 10^{-7}$ &
$10^{-4},10^{-6}$ & $10^{-4},2\times 10^{-7}$ 
\\ 
\hline 
$\mu,\mu'$ & 0.1,0.27 & 0.1,0.26& 0.22,0.45 & 0.1,0.35
\\ \hline
$s_1(\mu_s)/\eta$ & 0.958 & 0.969 & 0.821 & 0.922\\ \hline
$\mu_s$ &0.550&0.555&0.478&0.535\\ \hline
$R/R_{TAMKR}$ & $88.0\%$ & $92.0\%$ &57.3\% & 80.8\%
\end{tabular}
\end{table} 
\section{conclusion}
In conclusion, we have proposed an efficient and feasible decoy-state method to
do QKD over very lossy channel. The key rate for the main signal pulses
is around $57\%-92\%$ of the theoretically allowed maximal value. 
Our protocol uses vacuum and coherent states with intensities of $\mu,\mu',\mu_s$. All
coherent states can be used to distill the final key and $\mu_s$ is used as the main signal
pulses. As one may see from table I, the key rate of our protocol is rather close to 
that of the $ideal$
protocol. We believe the protocol shown here is the best choice among all existing decoy-state
protocols\cite{hwang,tot,lo4,my}. 
\acknowledgments
I am
 grateful to Prof. H. Imai for his long-term support. 
I thank Toshiyuki Shimono for his kindly help in the numerical calculation.

\end{document}